\documentclass[prd,twocolumn,aps,amsmath,nofootinbib,superscriptaddress]
{revtex4}

\usepackage{graphicx}
\usepackage{hyperref}
\usepackage{bm}
\usepackage{epsfig}
\usepackage{amsfonts}
\usepackage{mathrsfs}

\catcode`@=11
\let\savesort=\NAT@sort@cites
\newcommand\nosort[1]{\edef\NAT@cite@list{#1}}
\def\citenosort#1{\let\NAT@sort@cites=\nosort \cite{#1}%
   \let\NAT@sort@cites=\savesort}
\catcode`@=12 

\newcommand{\beq}{\begin{equation}}
\newcommand{\eeq}{\end{equation}}

\newcommand{\bea}{\begin{eqnarray}}
\newcommand{\eea}{\end{eqnarray}}

\newcommand{\comment}[1]{}

\begin{document}


\title{Negative vacuum energy densities and the causal diamond measure}

\author{Michael P.~Salem}
\affiliation{Institute of Cosmology, Department of Physics and Astronomy,\\
Tufts University, Medford, MA 02155}


\begin{abstract}
Arguably a major success of the landscape picture is the prediction of 
a small, non-zero vacuum energy density.  The details of this prediction 
depends in part on how the diverging spacetime volume of the multiverse 
is regulated, a question that remains unresolved.  One proposal, the 
causal diamond measure, has demonstrated many phenomenological successes, 
including predicting a distribution of positive vacuum energy densities 
in good agreement with observation.  In the string landscape, however, 
the vacuum energy density is expected to take positive and negative 
values.  We find the causal diamond measure gives a poor fit to 
observation in such a landscape --- in particular, 99.6\% of observers 
in galaxies seemingly just like ours measure a vacuum energy density 
smaller than we do, most of them measuring it to be negative.      
\end{abstract}


\maketitle

\section{Introduction}
\label{sec:introduction}

If our universe is one vacuum phase among many in an eternally inflating 
multiverse, our expectations for cosmological observables depend on the 
answers to three major questions:  (1) what is the distribution of the 
relevant physical parameters among generic vacuum states in the landscape, 
(2) how do conditional constraints associated with other characteristics 
of our universe, most notably anthropic constraints, affect this 
distribution, and (3) how do we add up / compare the distinct observations 
of the diverging number of observers.  The landscape prediction of the 
cosmological constant $\Lambda$~\cite{ccweinberg,Efstathiou,MSW,GLV,
Tegmark,VP,Peacock,ccbousso,ccsfc,ccprior}, made possible by our 
confidence in addressing questions (1) and (2) in this scenario, is 
arguable a major success of the multiverse picture.  Question (3) --- the 
so-called measure problem  --- remains unanswered, yet one may calculate 
expectations for $\Lambda$ given any specific proposal. 

One such proposal is the causal diamond measure~\cite{diamond,holo}, 
which regulates the diverging spacetime volume of the multiverse by 
restricting attention to only the spacetime volume of the causal 
horizon about a given worldline.  This measure has been shown to 
avoid the ``youngness paradox''~\cite{Tegmark:2004qd,BFY}, ``runaway 
inflation''~\cite{FHW,QGV,GS}, and under certain conditions ``Boltzmann 
brain domination''~\cite{Rees1,Albrecht,DKS02,Page1,Page2,Page06,BF06} 
--- three phenomenological pathologies in which the overwhelming majority 
of observers see a world in stark contrast with what we 
observe.\footnote{\label{f1}The scale-factor cutoff measure has also been 
shown to be free of these issues~\citenosort{Linde06,ccsfc,bbsfc,BFY2} 
(see also Refs.~\cite{GV08b,Bousso08} for measures closely related to the
scale-factor cutoff and causal diamond), while the stationary 
measure~\cite{Linde07,LVW} has been argued to be free of all but runaway 
inflation, some ideas to possibly get around which are suggested in 
Refs.~\citenosort{GarciaBellido:1994ci,QGV,Linde:2005yw,HWY}.}
The causal diamond measure has been combined with an entropy-counting
anthropic selection factor to successfully explain the observed value
of $\Lambda$, at least when attention is limited to positive 
$\Lambda$~\cite{ccbousso}, along with the observed values of other 
cosmological parameters~\cite{Cline:2007su}.  (See also 
Ref.~\cite{Omega}) It has also found phenomenological success using more 
explicit anthropic selection criteria~\cite{Bousso:2009ks}.  

On the other hand, it takes only one poor prediction to raise suspicion
on a given theory.  We calculate the full distribution (positive and 
negative values) of $\Lambda$, using the causal diamond measure and 
assuming a flat ``prior'' distribution among the very small values of 
$|\Lambda|$ in the landscape.  The result is in poor agreement with 
observation, with about 99.6\% of observers seeing a value smaller than 
what we measure.  This includes the effects of anthropic selection; in
fact compared to other calculations in the literature we use very 
restrictive selection criteria --- counting only galaxies with mass and 
virialization density very similar to those of the Milky Way --- to 
avoid counting hypothetical observers where possibly none in fact exist.  
Choosing very restrictive anthropic criteria can be viewed as simply 
further conditioning the predicted distribution, and therefore should 
only lead to increased accuracy~\cite{GV08}.  

One might be willing to accept our measurement of $\Lambda$ as a rare
statistical fluke, especially in light of any compelling theoretical
considerations and successful predictions of other cosmological 
observables, or if one believes that even our restrictive anthropic
criteria significantly overcount the number of observers measuring 
values of $\Lambda$ smaller than what we do.  Alternatively, given that 
we do not possess a firmly-grounded theoretical derivation of this or 
any other regulator of eternal inflation, one might see this result as 
providing guidance as to what directions to pursue with respect to 
formalizing other promising measures (see for example footnote~\ref{f1}) 
or modifying the causal diamond measure.

The remainder of this paper is organized as follows.  In 
Section~\ref{sec:background} we provide background to the problem, 
describing first our assumptions about the landscape, next the 
relevant aspects of the causal diamond measure, and finally our specific
anthropic criteria.  We perform the calculation in 
Section~\ref{sec:calculation}, where we also comment on some of the
uncertainties of the analysis.  A final discussion is given in 
Section~\ref{sec:discussion}.

\section{Background}
\label{sec:background}

\subsection{Landscape Assumptions}

String theory apparently possesses an enormous number of metastable
vacua, each with potentially different low-energy particle physics 
and/or vacuum energy densities~\cite{stringlandscape1,stringlandscape2,
stringlandscape3,stringlandscape4}.  We restrict attention to vacua
indistinguishable from our own except for the vacuum energy density, 
and also, as we explain below, to measurements of observers in galaxies 
much like the Milky Way.  In hindsight~\cite{ccweinberg,Efstathiou}, 
this limits the magnitude of the vacuum energy density to a range of 
values that is microscopic compared to the range of possibilities in 
the landscape.  For these reasons we assume the distribution of vacuum 
energy densities in relevant states of the landscape is essentially 
continuous (e.g. finer than the resolution of foreseeable observation) 
and flat,
\beq
I(\Lambda) = {\rm constant}\,.
\eeq      

Each of the various vacua of the landscape are physically realized 
during eternal inflation, during which spacetime regions in one de 
Sitter (dS) vacuum may tunnel to other dS or anti-de Sitter (AdS) 
vacua, forming new pocket universes (note that AdS vacua may not tunnel 
back to dS vacua).  This tunneling may be seen as proceeding through 
potential barriers between local minima (vacua) in the landscape.  We 
assume the various tunneling transition rates into a vacuum with very 
small $|\Lambda|$ are uncorrelated with the precise value $\Lambda$; 
this should be the case whenever the vacua with $\Lambda$ in any small 
interval $d\Lambda$ are surrounded by a diverse set of landscape 
potential barriers --- another property expected of an enormous 
landscape.  This conclusion has been demonstrated in several more simple 
landscape scenarios, see for example Refs.~\cite{SPV,S06,OS07,CSS07,S08}.

\subsection{Causal Diamond Measure}

The causal diamond measure~\cite{diamond,holo} regulates the diverging 
spacetime volume
of eternal inflation as follows.  One focuses on a single worldline, 
beginning in a given dS vacuum,\footnote{In general, the predictions
of the causal diamond measure depend on the choice of initial dS 
vacuum --- or how an ensemble of such vacua are weighted against each
other.  Although one might find it desirable that eternal inflation 
erase such dependence on initial conditions, there is at present no 
theoretical reason to demand that this be the case.} 
and considers the ensemble all possible future ``histories'' of that 
worldline.  All except a set of measure zero of the worldlines in this 
ensemble have finite duration, eventually terminating on an AdS 
singularity.  The multitude of possible histories in the above ensemble 
are weighted against each other according to their normalizable 
quantum-mechanical branching ratios.  For a given worldline in the 
ensemble, the causal diamond is constructed by finding the intersection 
of the future lightcone of the point at the beginning of the worldline, 
and the past lightcone of the point at which the worldline terminates.  
The fraction of observers measuring a given value of $\Lambda$ is then 
calculated by cataloging the full set of observers in the set of causal 
diamonds generated by the ensemble of worldline histories.

To proceed one can employ an important simplifying approximation.  First
note that vacua suitable to observers, which we call ``anthropic'' vacua, 
should be very rare among the full set of vacua in the landscape.  
Second, transitions from an anthropic dS vacuum to other dS vacua 
are suppressed by a factor $e^{-3/G\Lambda}$ relative to transitions to AdS
vacua, and meanwhile AdS vacua cannot transition back to dS vacua at all.  
These imply that among the worldline ``histories'' in the above ensemble, 
observers should overwhelmingly appear in just two situations:  either in 
dS vacua (with very small vacuum energies) that subsequently decay to AdS 
vacua (with relatively large-magnitude vacuum energy densities), in which 
case the worldline quickly terminates after decay to AdS, or in AdS vacua 
with small-magnitude vacuum energy densities, in which case the worldline 
terminates at the AdS singularity.  Therefore, to good approximation we 
can restrict attention to histories in the above ensemble in which only 
one anthropic vacuum is encountered, as either the second-last or the 
last vacuum before the worldline is terminated.

The causal diamond measure can then be conveniently divided into two
parts.  In the first part one assigns a ``prior'' 
probability $I$ to each type of pocket, according to the relative 
likelihood of the above worldline encountering such a pocket.  An example 
of such a calculation is made explicit in Ref.~\cite{diamond}; however for 
us the details are unimportant because, as described above, the prior 
distribution of $\Lambda$ is flat, $I(\Lambda)=$ constant.  Each anthropic 
pocket then receives an additional weight $A(\Lambda)$, corresponding to 
the typical number of observers measuring $\Lambda$ in the causal diamonds 
of the ensemble that include that anthropic pocket.  Since by hypothesis 
the worldline terminates soon after the decay of dS (or at the singularity 
of AdS), the typical intersection of a causal diamond with an anthropic 
pocket is simply the past lightcone of a random point on the hypersurface 
of decay (or the final singularity) of the anthropic 
pocket.\footnote{\label{lightcone}To
see this, note that the causal diamond stops growing, and begins to shrink, 
when half of the conformal time has elapsed between the start of the 
worldline and its end.  Meanwhile, the time rate of change of conformal 
time is one over the scale factor.  Since the scale factor grows 
exponentially during inflation, the pace of conformal time is exponentially 
suppressed at late times, the effect of which is the causal diamond almost 
always begins to shrink before a worldline reaches the reheating 
hypersurface of an anthropic vacuum.  
It might help to consider the following crude model:  the worldline begins
at the center of nucleation of an anthropic vacuum like ours, extends
through $N$ e-folds of exponential expansion at Hubble rate $H_I$ (modeling
inflation), and then immediately falls to exponential expansion at Hubble 
rate $H_\Lambda$ (modeling cosmological constant domination).  The ratio
of conformal time before the transition from $H_I$ to $H_\Lambda$ (crudely 
modeling reheating followed by radiation and matter domination) to that 
after this transition is roughly $(H_\Lambda/H_I)e^N$, which in anthropic 
vacua like ours is much larger than one unless the cosmological constant 
is exponentially suppressed relative to the value we measure.  Such small 
values of cosmological constant contribute negligibly to the total 
anthropic distribution.}
Henceforth we refer to this past lightcone as the causal diamond.  

Note that, as defined and implemented above, the ``causal diamond'' 
measure makes the same anthropic predictions as another form of ``causal
patch'' measure, which is defined exactly as above except with the role
of the causal diamond played by the past lightcone of the future endpoint 
of a given worldline.  (This equivalence is trivial, since as described 
above the intersection of the causal diamond with the anthropic pocket
is taken to be equal to the past lightcone in the anthropic pocket.)  Thus, 
the terms ``causal diamond'' and ``causal patch'' have been used 
interchangeably in the literature.  

Putting the above results together, we write
\beq
dP(\Lambda) \propto A(\Lambda)\,I(\Lambda)\, d\Lambda \,,
\label{P}
\eeq
where again $I(\Lambda)=$ constant.  The typical number of observers in 
the causal diamond can be written
\beq
A(\Lambda) \propto \int_0^{\tau_f}\! 
\rho_{\rm obs}(\Lambda,\tau)\, V_\diamond(\Lambda,\tau)\, d\tau \,,
\label{A}
\eeq   
where $\rho_{\rm obs}$ is the average number of observers per unit 
four-volume, $V_\diamond$ denotes the physical three-volume in the causal
diamond as a function of proper time $\tau$, and $\tau_f$ is the 
time of vacuum decay in an anthropic dS vacuum or the time of the 
final singularity in an anthropic AdS vacuum.  Defining conformal time
$\eta=\int d\tau/a(\tau)$ (where $a$ is the scale factor) and choosing 
the integration constant so that $\eta$ is negative and approaches zero 
as $\tau\to\tau_f$, we can write  
\beq
V_\diamond(\Lambda,\tau) \propto -a^3(\Lambda,\tau)\,
\eta^3(\Lambda,\tau) \,.
\label{causaldiamond}
\eeq

\subsection{Anthropic Selection}
\label{ssec:anthropic}

We seek a distribution of $\Lambda$ from which the value we measure can 
be considered as randomly drawn.  Although one might speculate that this 
applies to the (properly regulated) set of measurements made 
by all observers, it is less presumptuous to take our measurement as 
typical of those made by observers very much like us.  The only danger 
in specifying a more narrow notion of observer is the possibility that 
such observers exist for only a slim range of $\Lambda$, leaving little 
opportunity to falsify the prediction.  This will not be the case with 
our analysis.

We do not possess the technical skill to track the density of observers 
just like us.  One step to simply the problem is to restrict attention 
to a slice of the landscape on which all parameters except $\Lambda$ are 
fixed to the values we measure.  Thus we ask the conditional question: 
given what we know about the parametrization of our universe, what 
value of $\Lambda$ should we expect to measure?  

A second, more significant step to simplify the problem is to develop an 
astrophysical proxy for an observer. It is of course crucial that the 
set of proxies faithfully represents the set of observers.  We take as
our proxy ``Milky-Way like'' (MW) galaxies, by which we mean galaxies
constrained as much as is reasonably possible to resemble the Milky Way
(details are given below).  Unless our existence in a MW galaxy is itself 
atypical of observers like us --- a circumstance made unlikely by the 
fact that a diverse range of galaxies exist even in our universe --- this 
approach should only increase the accuracy of our prediction relative 
more general approaches.  Regardless, restricting attention to MW 
galaxies can simply be viewed as performing a more conditioned prediction.

To identify MW galaxies, we adopt the following simplified picture of
structure formation~\cite{PS}.  Consider first a comoving sphere enclosing 
a mass $M$, with density contrast $\sigma>0$.  The evolution of $\sigma$ 
can be analyzed using linear perturbation theory, or for instance by 
studying the evolution of a spherical top-hat overdensity.  The collapse 
density threshold $\delta_c$ is defined as the amplitude of $\sigma$, 
according to the linear analysis, when the spherical top-hat analysis says 
it has collapsed.  The collapse density threshold is not a constant when 
$\Lambda\ne 0$, but the variation is relatively small and to good 
approximation we can simply set $\delta_c=1.69$~\citenosort{PTV,ccsfc}.

In a more sophisticated analysis, the overdensity does not collapse to a 
point, but virializes as a halo with density $18\pi^2$ times the average 
cosmic matter density at that time it would have collapsed~\cite{Tegmark}.  
In fact, structure formation is hierarchical:  smaller comoving regions 
collapse and virialize first, and larger halos form as these accrete 
matter, merge, and `revirialize.'  In our universe the rate of such 
mergers decreases with time, furthermore the baryons cool and collapse 
into galaxies containing stars.  We approximate the full process of 
structure formation by identifying a critical time $\tau_*$, before which 
the baryons in a halo are continuously revirializing, and after which the 
evolution of the baryons depends only on the mass $M$ and the average 
density at $\tau_*$.  

We define a MW galaxy as one that has the same mass, virialization 
density, and age as the Milky Way, where by age we mean the time 
lapse $\Delta\tau$ between $\tau_*$ and when observers arise.  The time 
$\tau_*$ is different in different universes; it is found by solving 
$\rho_{\rm m}(\tau_*)=\overline{\rho}_*$, where $\rho_{\rm m}$ is the 
cosmic matter density and $\overline{\rho}_*$ is that when the Milky 
Way last virialized.  On the other hand $\Delta\tau$ should be the 
same across the set of universes that we consider; it is simply the
difference between the present cosmic time and that when the Milky 
Way last virialized.  

The distribution of $\sigma$ over comoving 
spheres enclosing mass $M$ is Gaussian with a standard deviation 
$\sigma_{\rm rms}(M,\tau)$ that depends on $M$ and grows with time.  
The probability that any such region collapses in a small interval
$d\tau$ about $\tau_*$ is therefore
\beq
dP_{\rm coll} \propto \frac{\dot{\sigma}_{\rm rms}}{\sigma_{\rm rms}^2}
\exp\left(\frac{1}{2}\frac{\delta_c^2}{\sigma_{\rm rms}^2}\right)
\bigg|_{M,\tau_*}d\tau \,,
\eeq
where the dot denotes differentiation with respect to $\tau$ (we 
have used that $\sigma/\sigma_{\rm rms}$ is constant in time 
according to the linear analysis).  Thus we write
\beq
\rho_{\rm obs}(\tau)\,d\tau \propto \delta(\tau-\tau_*-\Delta\tau)\,
\rho_{\rm m}(\tau)\, dP_{\rm coll} \,,
\label{obs1}
\eeq   
where the delta function arises because we have so constrained the 
set of observers, that they all arise at a single time in any given
universe.

There is another condition that should be considered.  Above we 
assumed that after some time $\tau_*$, the evolution of a galaxy is 
approximately independent of the surrounding cosmic environment.  
However this approximation must fail at least for late times and 
negative $\Lambda$, as ultimately such spacetimes collapse to a 
singularity.  To account for this, we impose the additional 
constraint,
\beq
\tau_* + \Delta\tau \leq \tau_f/2 \,,
\label{NLconst}
\eeq  
which ensures that we only count galaxies before any AdS vacua begin 
to collapse.

Note it is possible that Eq.~(\ref{NLconst}) does not go far enough.  
In universes with (positive and negative) values of $\Lambda$ smaller 
than ours, the rate of halo mergers and galaxy collisions (as a 
function of matter density) will be greater than in our universe.  
If these events are disruptive to the development of observers, for
instance by repeatedly resetting the process of star formation or by 
disrupting stable stellar systems with stellar fly-bys during galaxy 
collisions after star formation, then our calculation of 
$\rho_{\rm obs}$ overestimates the number of observers measuring 
$\Lambda$ smaller than we do.  A full consideration of this issue
appears to be rather formidable, and is not attempted in this work.  
Instead we make some basic observations about our results at the end 
of Section~\ref{sec:calculation}.

\boldmath
\section{Distribution of $\Lambda$}
\label{sec:calculation}
\unboldmath

Combining Eqs.~(\ref{P}--\ref{obs1}), the distribution of vacuum 
energy densities $\rho_\Lambda$ can be written

\beq
\frac{dP}{d\rho_\Lambda} \propto 
-\left[a^3\eta^3\rho_{\rm m}\right]_{\tau_*+\Delta\tau}
\left[\frac{\dot{\sigma}_{\rm rms}}{\sigma_{\rm rms}^2}
\exp\left(\frac{1}{2}\frac{\delta_c^2}{\sigma_{\rm rms}^2}\right)
\right]_{\tau_*} \,,
\label{P2}
\eeq
where again $a$ is the scale factor, $\eta$ is the conformal time, 
$\rho_{\rm m}$ is the matter density, $\sigma_{\rm rms}$ is the 
root-mean-square (rms) density contrast evaluated on a comoving scale
enclosing mass $M$, and $\delta_c=1.69$ is the collapse density
threshold.  The first term in brackets is proportional to the total
matter in the causal diamond when observers arise, a time $\Delta\tau$
after halo virialization at $\tau_*$, whereas the second term in 
brackets is proportional to the probability that a volume enclosing
mass $M$ will have virialized at time $\tau_*$.  One should also 
bear in mind we impose the constraint Eq.~(\ref{NLconst}).  We now
describe all of these pieces.  

As explained in Section~\ref{sec:background}, we restrict attention 
to pocket universes that are in every way like ours except for their 
vacuum energy densities.  In fact, since observers like us do not arise 
before matter domination, we can ignore the early radiation-dominated 
eras in these universes.  This allows for an analytic solution to the 
Einstein field equations, thus greatly simplifying the analysis.  The
cosmic matter density is then~\cite{ccbousso}
\begin{align}
\rho_{\rm m}(\tau) &= |\rho_\Lambda|\sin^{-2}
\left(\frac{3}{2}\frac{\tau}{\tau_\Lambda}\right)
\qquad \rho_\Lambda < 0 \displaybreak[0] \label{Nmatterdensity}\\
\rho_{\rm m}(\tau) &= \rho_\Lambda\sinh^{-2}
\left(\frac{3}{2}\frac{\tau}{\tau_\Lambda}\right) 
\qquad \rho_\Lambda \geq 0 \,,
\label{Pmatterdensity} 
\end{align}
where $\rho_\Lambda$ is the vacuum energy density and
\beq
\tau_\Lambda = \sqrt{3/8\pi G|\rho_\Lambda|} \,.
\eeq

As we restrict attention to pockets indistinguishable from ours except 
for the value of $\rho_\Lambda$, we normalize the scale factor so that 
at early times, it is independent of $\rho_\Lambda$.  Thus we write
\beq
a = (\rho_\Lambda^2/\rho_{\rm m})^{1/3} \,.
\eeq  
The conformal time is defined $\eta = \int d\tau/a(\tau)$.  We set the 
constant of integration so that $\eta$ is negative and approaches zero 
as $\tau\to\tau_f$.  As before, $\tau_f$ is the time of vacuum decay in 
dS vacua, in which case we can safely take $\tau_f\to\infty$, and 
$\tau_f$ is the time of the future singularity in AdS space, 
$\tau_f=(2\pi/3)\,\tau_\Lambda$.  This gives
\begin{widetext}
\begin{align}
\eta(\tau) &= -\left(\frac{3\tau_\Lambda}{8\pi G}\right)^{\!1/3}
\left\{\Delta\eta + \frac{2}{3}
\cos\left(\frac{3}{2}\frac{\tau}{\tau_\Lambda}\right)
\phantom{F}\!\!\!
_2{\rm F}_1\left[\frac{1}{2},\frac{5}{6},\,\frac{3}{2};\,
\cos^2\left(\frac{3}{2}\frac{\tau}{\tau_\Lambda}\right)\right]\right\}
& \rho_\Lambda &< 0 \displaybreak[0]\label{Nconformaltime}\\
\eta(\tau) &= -\left(\frac{3\tau_\Lambda}{8\pi G}\right)^{\!1/3}
\cosh^{-2/3}\left(\frac{3}{2}\frac{\tau}{\tau_\Lambda}\right)
\phantom{F}\!\!\!
_2{\rm F}_1\left[\frac{5}{6},\,\frac{1}{3},\,\frac{4}{3};\,
\cosh^{-2}\left(\frac{3}{2}\frac{\tau}{\tau_\Lambda}\right)\right] 
& \rho_\Lambda &\geq 0 \,,
\label{Pconformaltime}
\end{align}
\end{widetext}
where $\Delta\eta\simeq 2.429$ is simply an integration constant 
and $_2{\rm F}_1$ is the hypergeometric function.\footnote{As an 
anthropic AdS vacuum collapses, its radiation density will grow 
and eventually dominate the total energy density.  The appendix of 
Ref.~\cite{ccbousso} shows that the conformal time is changed 
negligibly by ignoring this radiation and instead tracking the 
matter density all the way to $a\to 0$.}    

The evolution of the rms density contrast $\sigma_{\rm rms}$ cannot
be written in closed form, even in our simplified model.  However, 
fitting functions accurate to the percent level are available in the
literature.  We use~\cite{Peacock}  
\beq
\sigma_{\rm rms}=\frac{3}{5}\sigma_{\rm eq}
\frac{(\rho_{\rm eq}/\rho_\Lambda)^{1/3}
\big[\textstyle{\frac{3}{2}} (\tau/\tau_\Lambda)\big]^{2/3}}
{\left[1+1.68\left(\tau/\tau_f\right)^{2.18}\right]\!
\left[1-4\left(\tau/\tau_f\right)^2\right]} \,,
\label{Nsigdef}
\eeq
for the case $\rho_\Lambda<0$, and~\cite{Tegmark} 
\beq
\sigma_{\rm rms}=\frac{3}{5}\sigma_{\rm eq}
\left[\left(\frac{\rho_{\rm m}}{\rho_{\rm eq}}\right)^{\!\!\alpha}\!\!
+\left(\frac{\rho_\Lambda}{G_\infty^3\rho_{\rm eq}}\right)^{\!\!\alpha}
\right]^{-1/3\alpha} , \,\,
\label{Psigdef}
\eeq
for the case $\rho_\Lambda \geq 0$, where $\alpha=159/200$ and 
$G_\infty=1.437$.  Here $\rho_{\rm eq}$ corresponds to the matter
density at matter-radiation equality (as measured in our universe).  
Of course our model does not contain radiation, but $\rho_{\rm eq}$ is
still a convenient reference to ensure our model matches onto the
observed evolution of $\sigma_{\rm rms}$.  Note that $\sigma_{\rm rms}$ 
near to and before matter-radiation equality differs significantly from 
that in Eqs.~(\ref{Nsigdef}) and~(\ref{Psigdef}), due to the presence 
of a non-growing mode, but this discrepancy is inconsequential since in 
both cases $\sigma_{\rm rms}$ is negligibly small at these times.

The time of last virialization $\tau_*$ is described in 
Section~\ref{ssec:anthropic} --- it corresponds to the time at which
MW galaxies must virialize in order to have the same virialization
density as the Milky Way, in our simple model.  If we take the Milky 
Way virialization time to be $\overline{\tau}_*$, then because the 
virialization density is proportional to the cosmic matter density
at the time of virialization,
\begin{align}
\tau_* &= \frac{2}{3}\tau_\Lambda \arcsin\left[
\sqrt{\frac{\rho_\Lambda}{\overline{\rho}_\Lambda}}
\sinh\left(\frac{3}{2}\frac{\overline{\tau}_*}
{\overline{\tau}_\Lambda}\right)\right] & \rho_\Lambda &< 0 
\label{taus1}\\
\tau_* &= \frac{2}{3}\tau_\Lambda {\rm arcsinh}\left[
\sqrt{\frac{\rho_\Lambda}{\overline{\rho}_\Lambda}}
\sinh\left(\frac{3}{2}\frac{\overline{\tau}_*}
{\overline{\tau}_\Lambda}\right)\right] & \rho_\Lambda &\geq 0 \,,
\label{taus2}
\end{align}  
where everywhere we use bars to denote quantities evaluated in our
universe.  Note also that the time lapse $\Delta\tau$, included in
our analysis to allow for the requisite evolution (including
planet formation and biological evolution) to bring the virialized
halo to the present state of the Milky Way, is simply 
\beq
\Delta\tau = \overline{\tau}_0 - \overline{\tau}_* \,,
\eeq
where $\overline{\tau}_0$ is the present age of our universe.
 
All that is left to determine Eq.~(\ref{P2}) is to fill in the 
various cosmological and astrophysical parameters.  We use WMAP-5
mean-value cosmological parameters~\cite{WMAP5}, along with 
CMBFAST~\cite{cmbfast} to determine the density contrast on a 
comoving scale enclosing $M=10^{12}$ solar masses,\footnote{For 
convenient reference we note the relevant cosmological parameter 
values are $\Omega_\Lambda=0.742$, $\Omega_m=0.258$, 
$\Omega_b=0.044$, $n_s=0.96$, $h=0.719$, and 
$\Delta_{\mathcal R}^2(k=0.02\,{\rm Mpc}^{-1})= 2.21\times 10^{-9}$.
These give $\sigma_{\rm rms}
(M=10^{12}M_\odot,\tau=\overline{\tau}_0)=2.03$ 
in our universe.} 
corresponding roughly to the mass scale of the Milky Way.  In 
fact, depending on the choice of $\Delta\tau$ one should choose a
somewhat smaller value of $M$, to account for accretion and minor
mergers during the interval $\Delta\tau$.  However our choice of 
$M$ is already a bit of an underestimate, and the primordial 
density contrast has a rather weak dependence on $M$.  The 
remaining astrophysical parameter is $\Delta\tau$.  Interestingly,
we find the results to be very insensitive to reasonable choices 
of $\Delta\tau$.  For the moment we choose 
$\Delta\tau=5\times 10^9$ years and later comment on the effect
of changing this.

\begin{figure}[t!]
\includegraphics[width=0.4\textwidth]{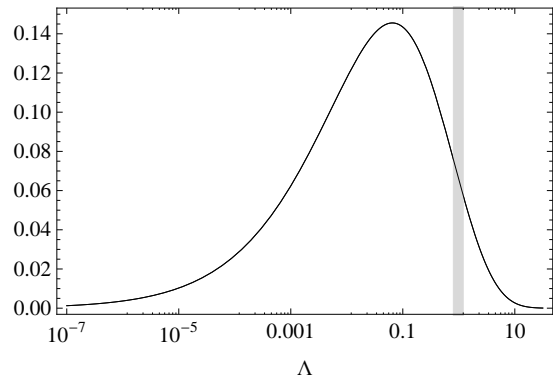} 
\caption{The normalized distribution of positive values of 
$\Lambda$, measured in units of the value we measure, 
$\overline{\Lambda}$, which is highlighted by the vertical bar.}
\label{fig1}
\end{figure}

We first compare to previous results.  Fig.~\ref{fig1} displays
the distribution of positive values of $\Lambda$ for the
parameter choices described above.  Although this distribution 
has not appeared before in the literature, Ref.~\cite{ccbousso}
calculated the distribution of positive values of $\Lambda$
using an entropy-counting approximation to implement anthropic 
selection, and restricting attention to the ``inner'' causal diamond, 
i.e. the intersection of what we have referred to as the causal 
diamond with the future lightcone of a point on the surface of 
reheating.  Comparing to Fig.~1 or Fig.~8 in Ref.~\cite{ccbousso}, we
see that our approach shifts the distribution to smaller values of 
$\Lambda$; however the curve still gives an acceptable fit to 
observation.\footnote{Although it is not evident from Fig.~\ref{fig1}, 
the distribution of $\Lambda$ diverges at $\Lambda=0$~\cite{HA}.  Yet 
the divergence is integrable over continuous distribution of 
$\Lambda$, so this is only a problem if observers can arise in vacua 
with $\Lambda$ exactly equal to zero.  As such vacua are expected to 
be supersymmetric, this is not the case for observers like us, and 
seemingly not the case for observers in general~\cite{Susskindbook}.}
Note that Fig.~\ref{fig1} is somewhat deceptive due to the long 
tail of the distribution toward small $\Lambda$.  In fact only 
about 6\% of observers measuring positive $\Lambda$ measure it 
to be larger than the value we measure.

The difference between our result and that of Ref.~\cite{ccbousso}
has two distinct sources: (1) Ref.~\cite{ccbousso} restricts attention 
to the above-mentioned inner causal diamond, whereas we use the full 
causal diamond (or, equivalently, the past lightcone, see 
footnote~\ref{lightcone} and surrounding discussion), and (2) 
Ref.~\cite{ccbousso} estimates the number of observers by integrating 
the entropy production, whereas we count the number of galaxies with 
mass, virialization density, and age equal to those of the Milky Way.
The choice of inner causal diamond was made in Ref.~\cite{ccbousso} to 
avoid counting the entropy produced at reheating, which would otherwise 
dominate the calculation~\cite{pbousso}.  Note however that this 
choice technically constitutes a different measure, the definition of 
which appears ad hoc.  For instance, the motivation to restrict to a 
causal patch is based on an analogy to black-hole 
complementarity~\cite{diamond,holo}; however there is clearly no 
problem with receiving information from beyond the inner causal diamond, 
as this occurs when we observe the cosmic microwave background.  
Meanwhile, to the extent that counting entropy production differs from 
counting MW galaxies, we consider the former to include ``observers'' 
who are rather unlike ourselves (or to omit observers who are much 
like ourselves), in which case the result is biased by observations 
from which it is more presumptuous for us to consider our measurement 
as randomly drawn.  To disentangle these effects for the interested
reader we note that, if one restricts to the inner causal diamond but 
otherwise adheres to our approach (equating observers with MW galaxies, 
etc.), one finds that about 11\% of observers measuring positive 
$\Lambda$ measure it to be larger than the value we measure.

\begin{figure}[t!]
\includegraphics[width=0.4\textwidth]{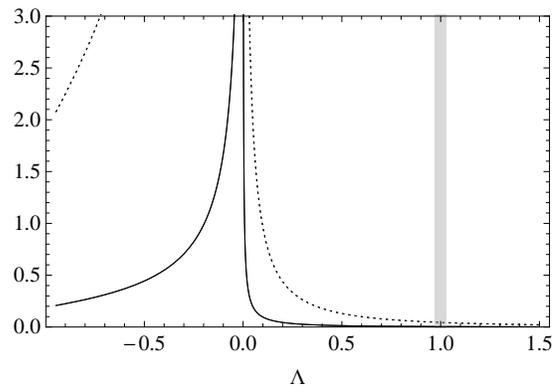} 
\caption{The normalized distribution of $\Lambda$ (solid), 
measured in units $\overline{\Lambda}$, which is highlighted 
by the vertical bar.  The dotted curve is the same distribution
but multiplied by a factor of ten for clarity.  The range of the
plot is chopped so as to implement the constraint of 
Eq.~(\ref{NLconst}).}
\label{fig2}
\end{figure}

The situation is much worse when we include negative values of
$\Lambda$.  The full distribution of $\Lambda$ is displayed in
Fig.~\ref{fig2}.  The range of the plot is chosen so as to chop the
distribution when the constraint of Eq.~(\ref{NLconst}) is violated.  
Due in large part to the relatively large number of observers who measure 
negative $\Lambda$, the fraction of observers measuring $\Lambda$ 
to be smaller than the value we measure is 99.6\% --- making our 
measurement appear as a roughly three standard deviation statistical 
fluke.  (If one restricts to the inner causal diamond, this fraction is 
98\%.)  The reason so many observers measure negative $\Lambda$ is 
easy to understand:  due to differences in global geometry, the 
comoving three-volume of the causal diamond on any constant-time 
hypersurface is much larger in AdS space than in dS space, for a 
given value of $|\Lambda|$.    

Because much of the problem is rooted in the relative volume-weighting 
of AdS and dS space, we do not expect the details of our attempt to 
identify Milky Way-like galaxies to significantly change the result.  
To illustrate this, consider two very different choices of 
$\Delta\tau$:  $\Delta\tau=0$ and $\Delta\tau=10^{10}$ years 
(for the moment we keep $M$ fixed).  Respectively, in these cases
we find 98.7\% and 99.7\% of observers measure a value of 
$\Lambda$ that is smaller than what we measure.  Actually, in the 
case of larger $\Delta\tau$, we should note that over such a time 
interval the mass of a typical galaxy grows significantly, so one 
might want to also decrease $M$.  Yet this is hardly helpful, due 
to the weak dependence of $\sigma_{\rm rms}$ on $M$.  If for instance
we choose $\Delta\tau=10^{10}$ years and $M=10^{11}$ solar masses, 
we still find 99.7\% of observers measure a value of $\Lambda$ 
less than what we measure.

Although our results are robust to choosing different time lapses 
$\Delta\tau$, we cannot exclude the possibility that we make a 
different type of error:  counting MW galaxies as if they have 
observers when in fact they do not.  As described at the end of
Section~\ref{ssec:anthropic}, this might be the case if halo
mergers and galaxy collisions --- which should occur at higher
rates in universes with smaller values of $\Lambda$ --- are 
detrimental to the development of observers.  A full analysis of
this issue appears rather formidable, so we instead make some 
basic observations.

First of all, if an increased rate of mergers keeps our basic
model of structure formation intact, but merely shifts the last
virialization time $\tau_*$ away from the values given by 
Eqs.~(\ref{taus1}--\ref{taus2}), then it is unlikely to 
significantly affect our results, for much the same reason that
adjusting $\Delta\tau$ does not significantly affect the results. 

\begin{figure}[t!]
\begin{tabular}{c}
\includegraphics[width=0.4\textwidth]{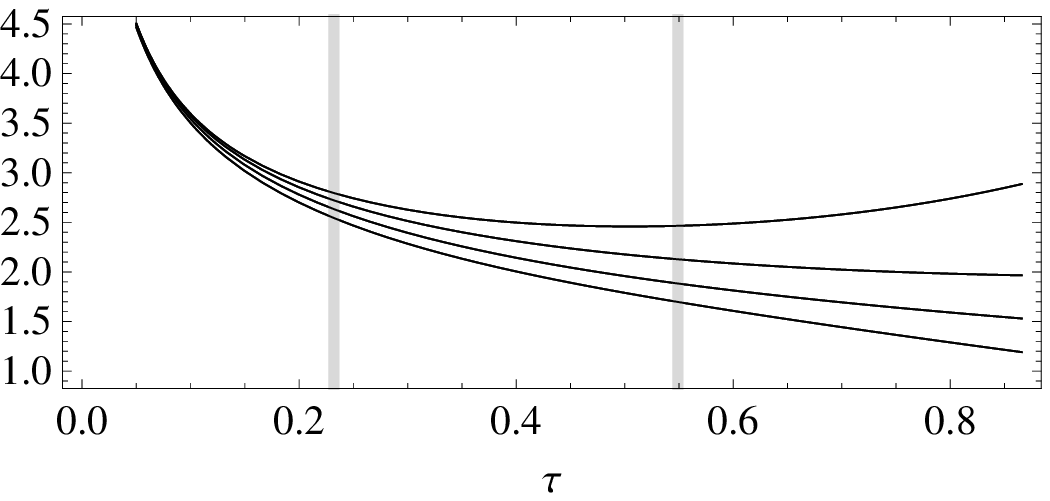} \\
\includegraphics[width=0.4\textwidth]{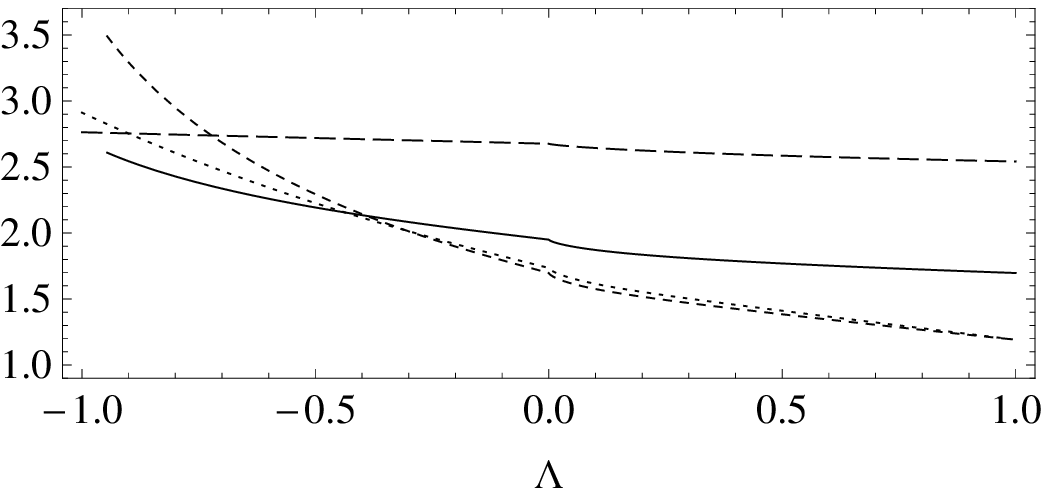}
\end{tabular}
\caption{Top: the time evolution of $\dot{\sigma}_{\rm rms}$, 
with $\tau$ measured in units of $\overline{\tau}_\Lambda$ 
($\overline{\tau}_0=0.87\overline{\tau}_\Lambda$), for 
$\Lambda=\overline{\Lambda}$ (bottom), 
$\Lambda=\overline{\Lambda}/4$ (next above),
$\Lambda=-\overline{\Lambda}/4$ (next above),
$\Lambda=-\overline{\Lambda}$ (top).  The vertical bars 
denote $\overline{\tau}_*(\Delta\tau=5\times 10^9\,{\rm yrs})$ 
and $\overline{\tau}_*(\Delta\tau=10^{10}\,{\rm yrs})$.  
Bottom: $\dot{\sigma}_{\rm rms}$ as a function of $\Lambda$,
measured in units of $\overline{\Lambda}$, evaluated at 
$\tau_*(\Delta\tau=5\times 10^9\,{\rm yrs})$ (solid),
$\tau_*(\Delta\tau=10^{10}\,{\rm yrs})$ (dashed),
$\tau_0(\Delta\tau=5\times 10^9\,{\rm yrs})$ (shorter dashed),
and $\tau_0(\Delta\tau=10^{10}\,{\rm yrs})$ (dotted). Recall 
that overlines indicate the values of quantities measured in
our universe.}
\label{fig3}
\end{figure}

Second, the merger rate itself should differ from one universe to 
another in a way closely related to the differing evolution of 
$\dot{\sigma}_{\rm rms}$.  Yet, because our anthropic constraints 
already limit us to a rather narrow range of $\Lambda$, the 
evolution of $\dot{\sigma}_{\rm rms}$ is not dramatically 
different across the range of anthropic conditions we consider.
This is illustrated in Fig.~\ref{fig3}.  The top panel plots the 
time evolution of $\dot{\sigma}_{\rm rms}$:  it is a decreasing 
function of time, except at late times in vacua with negative 
$\Lambda$, and at fixed times it is an increasing function of 
decreasing $\Lambda$.  The two vertical bars indicate the time
of last virialization, in our universe, for the two choices
$\Delta\tau=5\times 10^9$ years and $\Delta\tau=10^{10}$ years.

The bottom panel of Fig.~\ref{fig3} more clearly indicates the 
dependence on $\Lambda$ at two critical times in our model: the 
time of last virialization $\tau_*$ and the time of observation
$\tau_0$.  Note that the curve for 
$\tau_*(\Delta\tau=10^{10}\,{\rm yrs})$ is nearly constant --- 
this means selecting MW galaxies according to the corresponding
fixed virialization density is not much different than selecting
MW galaxies according to when $\dot{\sigma}_{\rm rms}$ falls 
below a certain critical level.  When the curve for $\tau_0$ is 
below the curve for $\tau_*$, this indicates that 
$\dot{\sigma}_{\rm rms}$ never rises above that critical level
(before the arrival of observers).  These curves might be taken
to indicate a stronger bound on small $\Lambda$ than our 
anthropic criteria assume; yet even if we set a lower bound as 
tight as $\Lambda\geq -\overline{\Lambda}/4$, then still 99.4\% 
of observers measure a value of $\Lambda$ smaller than we do 
(99.6\% for $\Delta\tau=10^{10}$ yrs). 

Finally, we note that because halo mergers and galaxy collisions
are probabilistic events, modest increases in the average rates
can be overcome by fortuitous circumstances.  In the context of
the above discussion, $\dot{\sigma}_{\rm rms}$ may be seen to 
give a measure of the average merger rate, but the actual merger
rate will vary from region to region depending on the local
properties of the density contrast.  Demanding fortuitous 
circumstances will introduce a statistical suppression factor 
when counting observers in vacua with smaller values of $\Lambda$, 
but if this factor does not change the order of magnitude of the 
density of observers, then the results would still be rather 
discouraging for the causal diamond measure.

\section{Discussion}
\label{sec:discussion}

We have calculated a distribution of positive and negative values
of $\Lambda$, using the causal patch measure, restricting attention
to universes otherwise like ours, and assuming a flat distribution 
of $\Lambda$ when $|\Lambda|$ is very small.  We found the value we 
measure to be roughly a three standard deviation outlier, with 
99.6\% of observers measuring values smaller than we do.

This result appears to be rather robust.  Our calculation used 
restrictive anthropic criteria, attempting to avoid mistakenly 
counting hypothetical observers where none in fact would exist.  
Our restrictive anthropic criteria might have missed observers, but 
these observers would live in galaxies that look rather different 
than ours.  Ignoring such observers is equivalent to simply further 
conditioning the distribution that we calculate.  We explored 
adjusting the parameters by which we attempt to identify Milky 
Way-like galaxies (these parameters are, to a large degree, 
uncertain), but found the results were insensitive to these 
changes.  Yet it is possible that even our restrictive anthropic 
criteria are not restrictive enough --- for example if we 
underestimate the effects of the increased numbers of mergers for 
smaller values of $\Lambda$.  Although we have not ruled out this
possibility, we provided some basic observations to suggest 
that this should not drastically change the results.

Still, depending on one's attitude, this sort of discrepancy on a single 
data point might not be seen as a significant problem.  One source 
of interest comes from the perspective that the causal diamond measure, 
though motivated by considerations of holography and black-hole 
complementarity, is not firmly grounded in fundamental theory (of 
course, neither are any of the other proposed regulators of eternal 
inflation).  If one presumes the ``correct'' measure gives a better
fit to observation, then our result might be seen as highlighting 
certain paths toward the correct measure.

For instance, it has recently been shown that the causal diamond
measure, given suitable initial conditions and restricting attention
to dS vacua, is equivalent to a seemingly very different approach 
based on an analogy to AdS/CFT complementarity~\cite{Bousso08}.  Yet
the approach of Ref.~\cite{Bousso08} appears to open up new 
possibilities for treating AdS and Minkwoski vacua in the multiverse.
Our result highlights the phenomenological importance that proposals
to resolve these ambiguities do not fall prey to `over-weighting' the
volume of AdS pockets.

Alternatively, one might find motivation to pursue other spacetime
measures.  Let us here note that in a Bayesian analysis that compares 
one measure to another, one would not compare the goodness-of-fits, 
but instead the probability that each measure assigns to $\Lambda$ 
being in a small interval $d\Lambda$ about the value we measure.  The 
anthropic criteria used in this paper are easily translated to a 
measure that weights fixed comoving thermalized volumes equally (such 
a measure suffers from Boltzmann brain domination, but nevertheless 
may serve as a reference).  In such a measure the probability assigned 
to the value of $\Lambda$ we observe is eight times that of the causal 
diamond measure.  Comparison to the ``no collapse'' scale-factor 
cutoff measure~\cite{ccsfc,BFY2} is somewhat more involved. We have 
checked and the scale-factor cutoff assigns about 32 times the 
probability of the causal diamond measure to the value of $\Lambda$ we 
measure.  (Both of these measures were shown to provide a good fit to
the observed value of $\Lambda$ in for instance Ref.~\cite{ccsfc}.)  Of 
course how one uses these differences depends on one's theoretical priors.

\begin{acknowledgments}
The author thanks Raphael Bousso, Andrea De Simone, Ben Freivogel, 
Alan Guth, Stefan Leichenauer, Alex Vilenkin, and I-Sheng Yang for 
very helpful discussions.  This work was supported by the U.S. National 
Science Foundation under grant NSF 322.
\end{acknowledgments}



\begin{thebibliography}{99}

\bibitem{ccweinberg}
  S.~Weinberg,
  ``Anthropic Bound On The Cosmological Constant,''
  Phys.\ Rev.\ Lett.\ {\bf 59}, 2607 (1987).

\bibitem{Efstathiou}
G.~Efstathiou, MNRAS {\bf 274}, L73 (1995).

\bibitem{MSW}
  H.~Martel, P.~R.~Shapiro and S.~Weinberg,
  ``Likely Values of the Cosmological Constant,''
  Astrophys.\ J.\ {\bf 492}, 29 (1998) [arXiv:astro-ph/9701099].

\bibitem{GLV}
  J.~Garriga, M.~Livio and A.~Vilenkin,
  ``The cosmological constant and the time of its dominance,''
  Phys.\ Rev.\ D {\bf 61}, 023503 (1999) [arXiv:astro-ph/9906210].

\bibitem{Tegmark}
  M.~Tegmark, A.~Aguirre, M.~J.~Rees and F.~Wilczek,
  ``Dimensionless constants, cosmology and other dark matters,''
  Phys.\ Rev.\ D {\bf 73}, 023505 (2006) [arXiv:astro-ph/0511774].

\bibitem{VP}
L.~Pogosian and A.~Vilenkin,
  ``Anthropic predictions for vacuum energy and neutrino masses in the  light
  of WMAP-3,''
  JCAP {\bf 0701}, 025 (2007) [arXiv:astro-ph/0611573].

\bibitem{Peacock}
  J.~A.~Peacock,
  ``Testing anthropic predictions for Lambda and the CMB temperature,''
  Mon.\ Not.\ Roy.\ Astron.\ Soc.\ {\bf 379}, 1067 (2007)
  [arXiv:0705.0898 [astro-ph]].

\bibitem{ccbousso}
  R.~Bousso, R.~Harnik, G.~D.~Kribs and G.~Perez,
  ``Predicting the Cosmological Constant from the Causal Entropic Principle,''
  Phys.\ Rev.\ D {\bf 76}, 043513 (2007) [arXiv:hep-th/0702115].

\bibitem{ccsfc}
  A.~De Simone, A.~H.~Guth, M.~P.~Salem and A.~Vilenkin,
  ``Predicting the cosmological constant with the scale-factor cutoff
  measure,''
  arXiv:0805.2173 [hep-th].

\bibitem{ccprior}
For earlier discussion of this idea, see
  A.~D.~Linde,
  ``The Inflationary Universe,''
  Rept.\ Prog.\ Phys.\  {\bf 47}, 925 (1984);
  T.~Banks,
  ``T C P, Quantum Gravity, The Cosmological Constant And All That..,''
  Nucl.\ Phys.\  B {\bf 249}, 332 (1985);
  J.~D.~Barrow and F.~J.~Tipler, 
  ``The Anthropic Cosmological Principle,''
  Oxford University Press (1986);
  A.~D.~Linde, in {\em Three hundred years of gravitation}, ed. by
  S.~W.~Hawking and W.~Israel, Cambridge University Press (1987).

\bibitem{diamond}
  R.~Bousso,
  ``Holographic probabilities in eternal inflation,''
  Phys.\ Rev.\ Lett.\  {\bf 97}, 191302 (2006)
  [arXiv:hep-th/0605263].

\bibitem{holo}
  R.~Bousso, B.~Freivogel and I.~S.~Yang,
  ``Eternal inflation: The inside story,''
  Phys.\ Rev.\  D {\bf 74}, 103516 (2006)
  [arXiv:hep-th/0606114].

\bibitem{Tegmark:2004qd}
M.~Tegmark,
``What does inflation really predict?,''
JCAP {\bf 0504}, 001 (2005)
[arXiv:astro-ph/0410281].

\bibitem{BFY}
  R.~Bousso, B.~Freivogel and I.~S.~Yang,
  ``Boltzmann babies in the proper time measure,''
  arXiv:0712.3324 [hep-th].

\bibitem{FHW}
  B.~Feldstein, L.~J.~Hall and T.~Watari,
  ``Density perturbations and the cosmological constant from inflationary
  landscapes,''
  Phys.\ Rev.\ D {\bf 72}, 123506 (2005)
  [arXiv:hep-th/0506235];

\bibitem{QGV}
  J.~Garriga and A.~Vilenkin,
  ``Anthropic prediction for Lambda and the Q catastrophe,''
  Prog.\ Theor.\ Phys.\ Suppl.\  {\bf 163}, 245 (2006)
  [arXiv:hep-th/0508005].

\bibitem{GS}
  M.~L.~Graesser and M.~P.~Salem,
  ``The scale of gravity and the cosmological constant within a landscape,''
  Phys.\ Rev.\  D {\bf 76}, 043506 (2007)
  [arXiv:astro-ph/0611694].

\bibitem{Rees1}
M.~Rees, {\it Before the Beginning}, p.221 (Addison-Wesley, 1997).

\bibitem{Albrecht}
A.~Albrecht and L.~Sorbo,
``Can the universe afford inflation?,''
Phys.\ Rev.\  D {\bf 70}, 063528 (2004)
[arXiv:hep-th/0405270];
see also
A.~Albrecht, 
``Cosmic inflation and the arrow of time,''
in {\it Science and Ultimate Reality}, ed. by
J.~D.~Barrow, P.~C.~W.~Davies, and C.~L.~Harper (Cambridge University
Press, 2003) [arXiv:astro-ph/0210527].

\bibitem{DKS02}
L.~Dyson, M.~Kleban and L.~Susskind,
``Disturbing implications of a cosmological constant,''
JHEP {\bf 0210}, 011 (2002)
[arXiv:hep-th/0208013].

\bibitem{Page1}
D.~N.~Page,
``The lifetime of the universe,''
J.\ Korean Phys.\ Soc.\  {\bf 49}, 711 (2006)
[arXiv:hep-th/0510003].

\bibitem{Page2}
D.~N.~Page,
``Susskind's challenge to the Hartle-Hawking no-boundary proposal and
possible resolutions,''
JCAP {\bf 0701}, 004 (2007)
[arXiv:hep-th/0610199].

\bibitem{Page06}
D.~N.~Page,
``Is our universe likely to decay within 20 billion years?,''
arXiv:hep-th/0610079.

\bibitem{BF06}
R.~Bousso and B.~Freivogel,
``A paradox in the global description of the multiverse,''
JHEP {\bf 0706}, 018 (2007)
[arXiv:hep-th/0610132].

\bibitem{Linde06}
A.~Linde,
``Sinks in the Landscape, Boltzmann Brains, and the Cosmological Constant
Problem,''
JCAP {\bf 0701}, 022 (2007)
[arXiv:hep-th/0611043].

\bibitem{bbsfc}
  A.~De Simone, A.~H.~Guth, A.~Linde, M.~Noorbala, M.~P.~Salem and A.~Vilenkin,
  ``Boltzmann brains and the scale-factor cutoff measure of the multiverse,''
  arXiv:0808.3778 [hep-th].

\bibitem{BFY2}
  R.~Bousso, B.~Freivogel and I.~S.~Yang,
  ``Properties of the scale factor measure,''
  arXiv:0808.3770 [hep-th].

\bibitem{GV08b}
  J.~Garriga and A.~Vilenkin,
  ``Holographic Multiverse,''
  arXiv:0809.4257 [hep-th].

\bibitem{Bousso08}
  R.~Bousso,
  ``Complementarity in the Multiverse,''
  arXiv:0901.4806 [hep-th].

\bibitem{Linde07}
A.~Linde,
``Towards a gauge invariant volume-weighted probability measure for
 eternal inflation,''
JCAP {\bf 0706}, 017 (2007)
[arXiv:0705.1160 [hep-th]].

\bibitem{LVW}
  A.~Linde, V.~Vanchurin and S.~Winitzki,
  ``Stationary Measure in the Multiverse,''
  JCAP {\bf 0901}, 031 (2009)
  [arXiv:0812.0005 [hep-th]].

\bibitem{GarciaBellido:1994ci}
J.~Garcia-Bellido and A.~D.~Linde,
``Stationarity of inflation and predictions of quantum cosmology,''
Phys.\ Rev.\  D {\bf 51}, 429 (1995)
[arXiv:hep-th/9408023].

\bibitem{Linde:2005yw}
A.~Linde and V.~Mukhanov,
``The curvaton web,''
JCAP {\bf 0604}, 009 (2006)
[arXiv:astro-ph/0511736].

\bibitem{HWY}
L.~J.~Hall, T.~Watari and T.~T.~Yanagida,
``Taming the runaway problem of inflationary landscapes,''
Phys.\ Rev.\  D {\bf 73}, 103502 (2006)
[arXiv:hep-th/0601028].

\bibitem{Cline:2007su}
  J.~M.~Cline, A.~R.~Frey and G.~Holder,
  ``Predictions of the causal entropic principle for environmental conditions
  of the universe,''
  Phys.\ Rev.\  D {\bf 77}, 063520 (2008)
  [arXiv:0709.4443 [hep-th]].

\bibitem{Omega}
  B.~Bozek, A.~J.~Albrecht and D.~Phillips,
  ``Curvature Constraints from the Causal Entropic Principle,''
  arXiv:0902.1171 [astro-ph.CO].

\bibitem{Bousso:2009ks}
  R.~Bousso, L.~J.~Hall and Y.~Nomura,
  ``Multiverse Understanding of Cosmological Coincidences,''
  arXiv:0902.2263 [hep-th].

\bibitem{GV08}
  J.~Garriga and A.~Vilenkin,
  ``Prediction and explanation in the multiverse,''
  Phys.\ Rev.\  D {\bf 77}, 043526 (2008)
  [arXiv:0711.2559 [hep-th]].

\bibitem{stringlandscape1}
  R.~Bousso and J.~Polchinski,
  ``Quantization of four-form fluxes and dynamical neutralization of the
  cosmological constant,''
  JHEP {\bf 0006}, 006 (2000)
  [arXiv:hep-th/0004134].

\bibitem{stringlandscape2}
  S.~Kachru, R.~Kallosh, A.~Linde and S.~P.~Trivedi,
  ``De Sitter vacua in string theory,''
  Phys.\ Rev.\  D {\bf 68}, 046005 (2003)
  [arXiv:hep-th/0301240].

\bibitem{stringlandscape3}
  L.~Susskind,
  ``The anthropic landscape of string theory,''
  arXiv:hep-th/0302219.

\bibitem{stringlandscape4}
  M.~R.~Douglas,
  ``The statistics of string/M theory vacua,''
  JHEP {\bf 0305}, 046 (2003)
  [arXiv:hep-th/0303194].

\bibitem{SPV}
  D.~Schwartz-Perlov and A.~Vilenkin,
  ``Probabilities in the Bousso-Polchinski multiverse,''
  JCAP {\bf 0606}, 010 (2006)
  [arXiv:hep-th/0601162].

\bibitem{S06}
  D.~Schwartz-Perlov,
  ``Probabilities in the Arkani-Hamed-Dimopolous-Kachru landscape,''
  J.\ Phys.\ A  {\bf 40}, 7363 (2007)
  [arXiv:hep-th/0611237].

\bibitem{OS07}
  K.~D.~Olum and D.~Schwartz-Perlov,
  ``Anthropic prediction in a large toy landscape,''
  JCAP {\bf 0710}, 010 (2007)
  [arXiv:0705.2562 [hep-th]].

\bibitem{CSS07}
  T.~Clifton, S.~Shenker and N.~Sivanandam,
  ``Volume Weighted Measures of Eternal Inflation in the Bousso-Polchinski
  Landscape,''
  JHEP {\bf 0709}, 034 (2007)
  [arXiv:0706.3201 [hep-th]].

\bibitem{S08}
  D.~Schwartz-Perlov,
  ``Anthropic prediction for a large multi-jump landscape,''
  JCAP {\bf 0810}, 009 (2008)
  [arXiv:0805.3549 [hep-th]].

\bibitem{PS}
  W.~H.~Press and P.~Schechter,
  ``Formation of galaxies and clusters of galaxies by selfsimilar gravitational
  condensation,''
  Astrophys.\ J.\  {\bf 187} (1974) 425;
  J.~M.~Bardeen, J.~R.~Bond, N.~Kaiser and A.~S.~Szalay,
  ``The Statistics Of Peaks Of Gaussian Random Fields,''
  Astrophys.\ J.\  {\bf 304}, 15 (1986).

\bibitem{PTV}
  M.~Tegmark, A.~Vilenkin and L.~Pogosian,
  ``Anthropic predictions for neutrino masses,''
  Phys.\ Rev.\  D {\bf 71}, 103523 (2005)
  [arXiv:astro-ph/0304536].

\bibitem{WMAP5}
  J.~Dunkley {\it et al.}  [WMAP Collaboration],
  ``Five-Year Wilkinson Microwave Anisotropy Probe (WMAP) Observations:
  Likelihoods and Parameters from the WMAP data,''
  arXiv:0803.0586 [astro-ph];
  E.~Komatsu {\it et al.}  [WMAP Collaboration],
  ``Five-Year Wilkinson Microwave Anisotropy Probe (WMAP)
  Observations:  Cosmological Interpretation,''
  arXiv:0803.0547 [astro-ph].

\bibitem{cmbfast}
  U.~Seljak and M.~Zaldarriaga,
  ``A Line of Sight Approach to Cosmic Microwave Background Anisotropies,''
  Astrophys.\ J.\  {\bf 469}, 437 (1996)
  [arXiv:astro-ph/9603033].

\bibitem{HA}
  L.~Mersini-Houghton and F.~C.~Adams,
  ``Limitations of anthropic predictions for the cosmological constant
  $\Lambda$: Cosmic Heat Death of Anthropic Observers,''
  Class.\ Quant.\ Grav.\  {\bf 25}, 165002 (2008)
  [arXiv:0810.4914 [gr-qc]].

\bibitem{Susskindbook}
L.~Susskind, {\it The Cosmic Landscape}, 
Little, Brown, and Company, New York (2005).

\bibitem{pbousso}
R.~Bousso, private communication.


\end{thebibliography}
\end{document}